\documentclass[twocolumn,prb,showkeys,showpacs,floatfix]{revtex4}
\usepackage[dvips]{graphicx}
\usepackage{dcolumn}
\usepackage{amsmath}
\usepackage{bm}

\begin{document}

\title{Electronic structure of crystalline  binary and ternary Cd-Te-O compounds}
\author{E. \surname{Men\'{e}ndez-Proupin}}
\email{eariel99@yahoo.com}
\altaffiliation[On leave from ]{University of Havana, Cuba}
\altaffiliation[Present address: ]{University of Chile, Faculty of Science, Las Palmeras
 3425, {\~N}u{\~n}oa, Santiago, Chile.}
\author{G. Guti\'errez}
\altaffiliation[Also at ]{Condensed Matter Theory Group, Uppsala University, Sweden.}
\affiliation{Departamento de F\'\i sica, Universidad de Santiago de Chile,
Casilla 307, Santiago 2, Chile}
\author{E. Palmero}
\affiliation{Facultad de F{\'\i}sica-IMRE, Universidad de La Habana, San L\'azaro y L,
Vedado 10400, La Habana, Cuba}
\altaffiliation[Present address: ]{Centro de Investigaciones del Petr\'oleo, Havana, Cuba.}
\author{J.~L.~\surname{Pe\~na}}
\affiliation{Departamento de F{\'\i}sica Aplicada, CINVESTAV-IPN Unidad M\'erida,
A.~P.~73 Cordemex, M\'erida, Yucatan CP~97130, M\'exico}
\date{\today}

\begin{abstract}
The electronic structure of crystalline CdTe, CdO, $\alpha$-TeO$_2$,
CdTeO$_3$ and Cd$_3$TeO$_6$ is studied by means of first principles calculations.
The band structure, total and partial density of states, and charge densities are presented.
For $\alpha$-TeO$_2$ and CdTeO$_3$, 
Density Functional Theory within the Local Density Approximation (LDA)
correctly describes the insulating character of these compounds. 
In the first four compounds,
LDA underestimates the 
optical bandgap by roughly 1 eV. Based on this trend, we predict an optical 
bandgap of 1.7 eV for Cd$_3$TeO$_6$. This material shows an isolated conduction 
band with a low effective mass, thus explaining its semiconducting character observed recently.
In all these oxides, 
the top valence bands are formed mainly from the O 2p electrons. On the other hand,  
the binding energy of the Cd 4d band, relative to
the valence band maximum, in the ternary compounds is smaller than in CdTe and CdO.
\end{abstract}
\maketitle

\section{Introduction}

Cadmium Telluride, CdTe,  has been considered a prototype of II-VI semiconducting compounds for more than 30 years.
It has found important applications in
 $\gamma$-ray detectors, infrared windows, solar cells and other optoelectronic
 devices.\cite{zanio78}
 Having being the object of many of the early studies in semiconductor science, the difficulties
 found in obtaining high quality crystals hampered its research and
 applications in cutting edge electronic technology. Recently CdTe
  has received renewed interest due to the search for cheap technologies
 for mass production of solar cells that do not require high quality monocrystals.
 The fact that the
 fundamental absorption gap of CdTe lies in the region of maximum intensity of solar radiation makes
 it an important material for solar energy conversion on the earth. The solar cells based on heterojunctions of CdTe/CdS are a valuable option.

 The development of a complete technology requires complementary materials that could be used for substrates, coats, contacts, gates, etc. Among these materials
 are the amorphous CdTe oxides
 obtained by r.f. sputtering, which have been
 extensively studied by optical spectroscopy,  X-ray photoemission spectroscopy and Auger
 electron spectroscopy.
\cite{espinoza91,espinoza93,zapata94,zapata97,elazhari97,
bartolo99,caballero98,iribarren99,%
 arizpe00,iribarren01,bartolo02,bartolo02b}
 These alloys are transparent  insulators whose optical gap can be tuned according to
 the content of oxygen between 1.5 and 4 eV, and
 can play a role in CdTe technology similar to that of SiO$_2$ in Si technology.
  Native oxides have also been identified on CdTe
  surfaces. \cite{ebina80,davis81,wang87,choi88,heiba03}
  However, given the difficulties produced by its amorphous structure,
 the structural and electronic properties of these materials are mostly unknown.
 Even the nature of oxides in CdTe surfaces is still a subject of controversy.
 \cite{miotto03}

 The development of a thorough  understanding of the electronic properties  of the
 amorphous CdTe oxides require a careful study of the crystalline compounds of Cd, Te, and O. These compounds provide information about short range order and local electronic
 properties that are determinant in the
 amorphous regime.
 The Cd-Te-O compounds can also provide valuable information to help understand
 the processes of oxygen  interaction with crystalline CdTe, including adsorption and diffusion.

Modern ab-initio simulation provides a powerful tool to study complex
compounds and to predict structural properties with reasonable confidence.
Although there are many reports of ab-initio calculations of Cd and Te compounds,
including c-CdTe,\cite{wei88,vogel95,vogel96,albrecht97} l-CdTe,\cite{chelikowsky98}
c-CdO\cite{boettger83,jaffe91,vogel96,dou98,wang01cdo,guerrero02}and TeO$_n$
clusters\cite{kowada96,uchino96,suehara94,suehara95,berthereau96,suehara98,yakovlev02},
the crystalline TeO$_2$ and the ternary compounds remain largely unexplored.
Only one ab-initio study of a Cd-Te-O system, which addresses the O adsorption on
CdTe surfaces, is known to the  authors.\cite{miotto03}
Despite being a profitable material for electromechanical and acoustooptical devices,\cite{yakovlev02} in the $\alpha$ phase (paratellurite)
even crystalline TeO$_2$
has never been the object of a full electronic structure calculation.

The main difficulty for semi-empirical or ab initio simulation of Cd-Te-O compounds
has been the large number of  atoms per unit cell in the known phases,
\cite{thomas88,landoltIII7b,burckhardt82,kraemer85} (e.g. 12 atoms
in $\alpha$-TeO$_2$, 24 in $\beta$-TeO$_2$,
20 in Cd$_3$TeO$_6$,  40 in CdTeO$_3$,
224 in CdTe$_2$O$_5$). Moreover Cd atoms have ten electrons in level 4d
which lie in the range of the valence band of Cd compounds and increase significantly
the computer resources necessary for the calculations.
In certain cases, the effect of the Cd 4d electrons can be
represented with the help of  pseudopotentials plus a partial core correction scheme;
 this approach has been successful for molecular dynamic simulations.\cite{chelikowsky98}
However,
every use of this scheme needs to be tested, as the Cd 4d electrons are known
to influence several important properties, e.g. the structure of the valence band, bandgaps,
cohesive energy, and lattice parameter.\cite{wei88}

 In this paper we report the first ever calculation of the electronic properties
of crystalline $\alpha$-TeO$_2$ (paratellurite), monoclinic CdTeO$_3$ (Cadmium oxotellurate(IV)), 
and Cd$_3$TeO$_6$ (Cadmium hexaoxotellurate). In order to analyze
 the nature of the chemical bond in these materials, and
 to evaluate the systematic errors inherent to the theory,
 we also analyze the electronic structure
 of the simpler semiconducting compounds CdTe and CdO.

\section{Structural properties}

\begin{figure}[!bh]
\includegraphics[width=4.0cm]{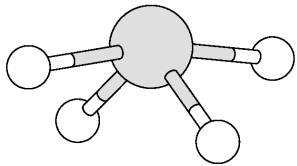}
\caption{Structural unit of $\alpha$-TeO$_2$. Generated with
XCrysSDen\protect\cite{xcrysden1,xcrysden2}.\label{fig:teo2-environment}}
\end{figure}

\begin{figure}[!tbh]
\includegraphics[width=6.5cm]{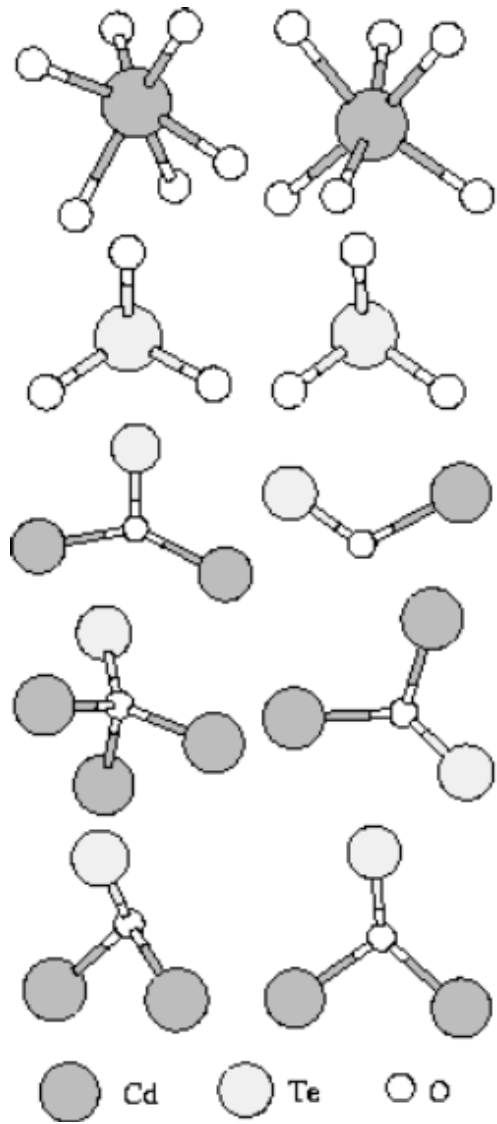}
\caption{Atomic environments in CdTeO$_3$. Generated with
XCrysSDen\protect\cite{xcrysden1,xcrysden2}.\label{fig:cdteo3-environment}}
\end{figure}

\begin{figure}[!tbh]
\includegraphics[width=6.5cm]{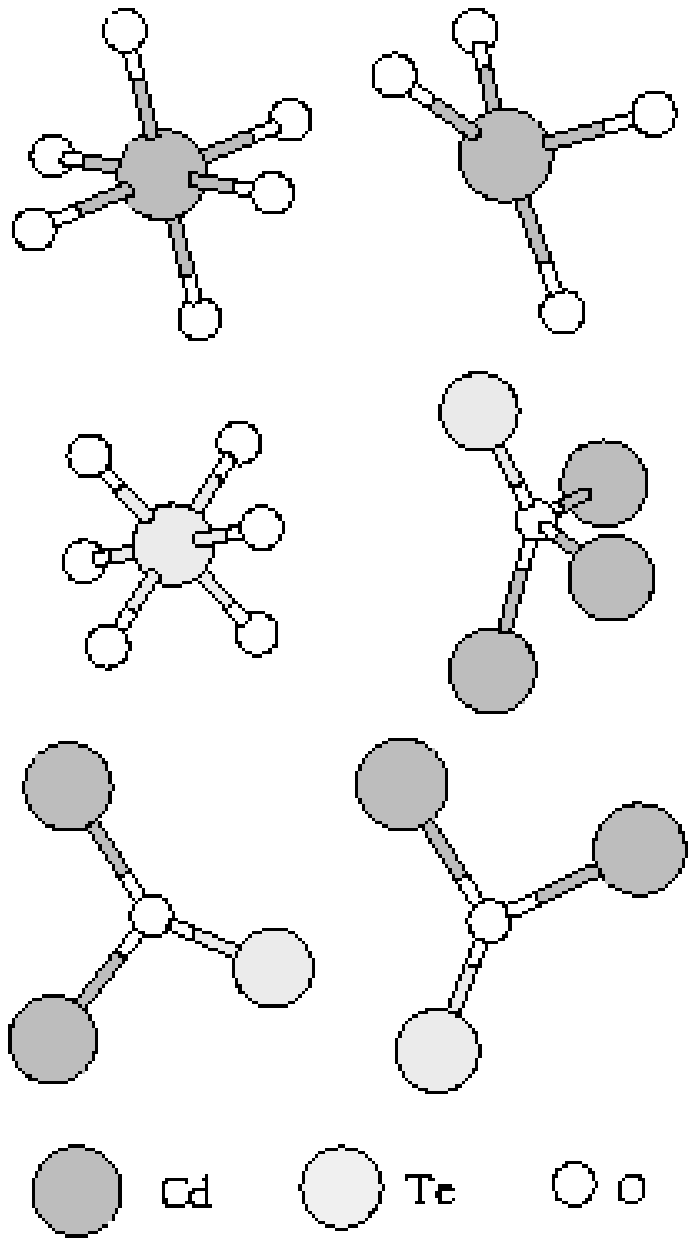}
\caption{Atomic environments in Cd$_3$TeO$_6$. Generated with
XCrysSDen\protect\cite{xcrysden1,xcrysden2}. \label{fig:cd3teo6-environment}}
\end{figure}

 TeO$_2$ is a mixed covalent-ionic insulator, the most abundant phase of which is
 $\alpha$-TeO$_2$ or paratellurite. Paratellurite structure
 has a tetragonal lattice with
 four formula units per unit cell. The space group has been identified as P$4_12_12$ and
 P$4_32_22$.\cite{ICSD} In this work we have studied the electronic properties for
 the most recently reported.\cite{thomas88}
At the short-range level, the structural units of paratellurite are TeO$_4$
trigonal bipyramids (tbp) with Te atoms at apex (see Fig. \ref{fig:teo2-environment}).
The corners of the tbps
are occupied by two equatorial oxygen atoms (O$^{eq}$), with Te-O$^{eq}$
bond length 1.88 \AA{} and angle O$^{eq}$-Te-O$^{eq}$ of 103$^{\circ}$,
and two axial oxygens (O$^{ax}$), with Te-O$^{ax}$
bond length 2.12 \AA{} and angle O$^{ax}$-Te-O$^{ax}$ of 168$^{\circ}$.
The tbps form a three dimensional network, interconnected through an equatorial and an axial oxygen.
This short-range order can be understood as produced by
Te sp$^3$ hybridization. This configuration consists of
the sharing of four of six valence electrons, and the repulsion of four
 bonds by the lone pair of Te electrons.\cite{mackay}
 Alternatively,
 the basic unit can be considered as a TeO$_6$ distorted octahedron
 (see Fig. \ref{fig:teo4tbp}), which is obtained by adding to a tbp the next nearest
 pair of O at 2.86 \AA{} from the central Te atom. From this point of view  the crystal structure
 of  paratellurite is considered a deformation of the structure of rutile. This approximation
  has been used in previous bandstructure calculations.\cite{robertson79,svane87}
  Ab initio calculations of TeO$_4$ and TeO$_6$
  clusters\cite{suehara94,suehara95,suehara98} show that the Te-O$^{ax}$ bond is less
  covalent than Te-O$^{eq}$, and the bond with the third pair of oxygen atoms is rather ionic. Recent
  studies\cite{mirgorodsky00,yakovlev02} show that the dynamic properties of paratellurite are
  better described if the basic unit is considered as molecular TeO$_2^{eq}$, while Te-O$^{ax}$
  bonds are understood to be intermolecular contacts.
These bonds are  important for the non linear
  dielectric properties.\cite{yakovlev02}

Crystalline CdTeO$_3$ has been obtained in a cubic phase and several monoclinic
phases.\cite{kraemer85,landoltIII7b} The cubic phase has also been reported in
oxidized CdTe.\cite{wang87,elazhari00,arizpe00,heiba03}
Only one of the monoclinic structures
has been totally determined.\cite{kraemer85} This structure belongs to the P2$_{1}/c$ monoclinic
crystallographic class, with the atoms occupying 10 non-equivalent positions.
Figure \ref{fig:cdteo3-environment} shows the first
coordination shell of all the non-equivalent atoms.
The structural units are distorted TeO$_3$ trigonal pyramids that are linked to each other with
Cd atoms. This arrangement is similar to
the one found in tellurite glasses.\cite{uchino96,sekiya92}
 There are Cd and Te atoms each of which occupy
two non-equivalent sites, and O
atoms in six  non-equivalent sites. Both Cd atoms are 6-fold coordinated with
Cd-O bond lengths in the 2.197-2.476~\AA{} range.
The bond angles deviate as much as 31 degrees 
from the ideal octahedral bond angles of 90 and 180 degrees. 
 O atoms always link one Te atom
 with one, two, or three Cd atoms, and the average coordination of O atoms is three.

 Cd$_3$TeO$_6$ has been recently investigated due to its
 semiconducting properties and has been
 proposed as a n-type thermoelectric material.\cite{shan02}
Its atomic structure\cite{burckhardt82} is described as a deformed perovskite-type with pseudo-orthorhombic
monoclinic space symmetry P2$_{1}/n$,
\footnote{P2$_{1}/n$ is a non-standard form of the space group 14, which is the
same as the standard P2$_{1}/c$, except that the lattice vectors $\mathbf{a}$
and $\mathbf{b}$ are chosen  in such a way that $\mathbf{a}+\mathbf{c}$
is parallel to the glide translation. \protect\cite{itc}} in which
the B-sites are occupied by Cd and Te, and the A-sites by Cd.
Adding the oxygen atoms, a total of 6 non-equivalent positions
 are occupied.  The local environments of
 the non-equivalent atoms are shown in Fig.~\ref{fig:cd3teo6-environment}.
 The first coordination shell of Te atoms is composed of three pairs of O atoms,
each pair at a bond length distance of
 1.904, 1.924,  and 1.948~\AA{}.  These lengths lie in the typical  range
 of paratellurite $\alpha$-TeO$_2$  and tellurite glasses.\cite{sekiya92,kowada96}
Cd atoms occupy two non-equivalent sites. One of
them is surrounded by three pairs of  O atoms
at bond lengths of  2.211, 2.311, and 2.350 \AA{} and bond angles near 90$^{\circ}$.
This arrangement is the same as in CdO with rocksalt structure, where the Cd-O bond length
 at room temperature is 2.344~\AA.\cite{landolt} 
The other Cd atom is surrounded by four oxygen atoms at
2.237, 2.251, 2.278, and 2.297 \AA. Three of these oxygen atoms lie in nearly the
same plane as the Cd atom, and the other one lies nearly perpendicular. This complex
configuration can be understood as originated by the electrostatic repulsion of the other
four O atoms located at 2.596, 2.736, 2.761, and 3.010 \AA{},
which complete a distorted octahedron, but are too far apart to be considered as
bound.
The O atoms occupy three non-equivalent crystallographic positions and are linked to one
Te atom and two or three Cd atoms. The two
arrangements of one O atom with two Cd atoms and one Te atom are nearly
planar.

The greatest difference between CdTeO$_3$ and Cd$_3$TeO$_6$  is that tellurium atom is
3-fold oxygen coordinated in the first and 6-fold coordinated in the latter.
This is
due to the fact that the content of tellurium in CdTeO$_3$ is twice that of  Cd$_3$TeO$_6$,
if  CdTeO$_3$ is regarded as Cd$_2$Te$_2$O$_6$. Moreover, the Te octahedral coordination
 is typical in compounds where Te has degree of oxidation VI.

In summary, in both CdTeO$_3$ and Cd$_3$TeO$_6$,
there are only Cd-O and Te-O bonds. Oxygen atoms
always bridge one tellurium atom with one, two
or three cadmium atoms. In the most typical cases an oxygen atom bridges one Te atom with two Cd atoms,
 with local quasi-planar arrangement.

\section{Electronic structure}

\begin{figure}
\includegraphics[width=8.0cm]{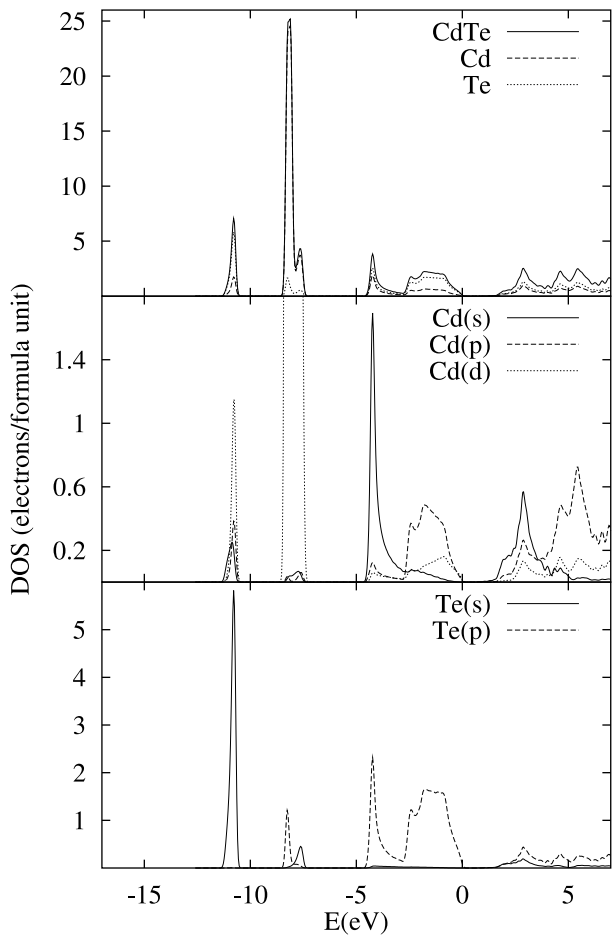}
\caption{Total and partial density of states of CdTe.}
\label{fig:pdoscdte}
\end{figure}

We have performed calculations of the electronic structure in the framework of the
Density Functional Theory (DFT)\cite{hk} with the Perdew-Zunger\cite{pz81}
parametrization of the
Local Density Approximation (LDA).
The calculations have been performed using the ab-initio total-energy program
VASP (Vienna ab-initio simulation program).\cite{vasp1,vasp2,vasp3,vasp4}
The valence electronic eigenfunctions are obtained from an expansion in plane
waves, while the core electrons are replaced by the ultrasoft pseudopotentials
supplied in the
VASP package. The cutoff of the plane wave expansion depends upon
the material composition. We have used the maximal values of the recommended
cutoffs of the components: 168, 115, and 396 eV for Cd, Te, and O, respectively.

 To obtain a smooth density of states (DOS),
  the first Brillouin zone has been
sampled with k-points generated according to the
Monkhorst-Pack scheme.
We used $N\times N \times N$ grids, with $N=29$ for CdTe, $N=33$ for CdO,
$N=10$  for TeO$_2$, $N=4$
 for CdTeO$_3$, and $N=6$ for Cd$_3$Te$O_6$. To eliminate unphysical oscillations
 in the density of states (DOS) due to the k-space grid, the DOS has been broadened
 by gaussian functions with 0.2 eV variance. However, for CdTe a broadening of  0.1 eV
 was used to  display the structure of the Cd 4d band.
 Analyzing the partial DOS we can  identify the nature of the peaks observed in the
 DOS, and correlate the features of the total DOS with the crystal structure.
 The partial DOS are obtained projecting the
 electron wave functions onto atomic-like orbitals, and integrating them
 in atom-centered spheres. The radii of these spheres are the atomic radii times
 a factor that equals the sum of the volumes of the atomic spheres to the
 cell volume. Our selection of the atomic radii is discussed in the Appendix.

As a check for the calculations we have performed optimizations of
the unit cell volume and
the ionic positions of CdTeO$_3$ and CdTe. For CdTeO$_3$,
the obtained lattice parameters were 4 \% smaller in the LDA than the experimental ones,
while the ionic displacements are smaller than 1 \% of the lattice parameters. For CdTe, the
theoretical lattice constant is 6.43 \AA{} in the LDA, and 6.62~\AA{} in the framework of the
Generalized Gradient Approximation (GGA), both being
within 2~\% of the experimental value 6.481~\AA.\cite{landolt}

As a complement to the  band diagram and the partial DOS analysis, we
use the electron localization function (ELF)\cite{becke90,savin92,kohout96,haussermann94}
to obtain information on interatomic bonds.
The ELF is a measure of the electron
localization in space, and helps to visualize the shell structure of atoms and solids.
The ELF is defined in terms of the  local  electronic
density $\rho(\mathbf{r})$, the excess of kinetic energy  due to the Pauli exclusion
principle $T(\mathbf{r})$, and the Thomas-Fermi kinetic energy density  $T_h(\mathbf{r})$.
\begin{equation}
\mathrm{ELF}=\left[ 1+\left(\frac{T(\mathbf{r})}{T_h(\mathbf{r})}\right)^2 \right]^{-1}.
\end{equation}
In a DFT calculation, $T(\mathbf{r})$ and $T_h(\mathbf{r})$ are evaluated from the
density and the Kohn-Sham orbitals $\varphi_i(\mathbf{r})$, as
\begin{eqnarray}
T(\mathbf{r}) &=&\frac{1}{2}\sum_{i} \vert\nabla\varphi_i(\mathbf{r})\vert^2
-\frac{1}{8}\frac{\vert\nabla\rho(\mathbf{r})\vert^2}{\rho(\mathbf{r})}    \\
T_h(\mathbf{r}) &=&0.3(3\pi^2)^{2/3}\rho(\mathbf{r})^{5/3}
\end{eqnarray}
By definition, ELF values range between 0 and 1. An ELF value
close to 1 at a given point indicates a high degree of localization of the electrons at
that point. For the homogeneous electron gas, which is used as a reference system
in its definition, the ELF is equal to 0.5.

\subsection{CdTe}

Cadmium telluride is a prototype of II-VI semiconductors.
It has the  zinc-blende structure, where the
anions arrange in a cubic compact structure, and the cations occupy one half of the tetrahedral sites
between the anions.  The band structure of
CdTe was determined 30 years ago\cite{chelikowsky73,chelikowsky76}
and has been obtained by ab initio calculation many
times (see e.g. Refs. \onlinecite{wei88,yeh94,vogel96,albrecht97,chelikowsky98}).
In Fig. \ref{fig:pdoscdte} we show the DOS.
The energies are referred to the valence band maximum.
The top valence band (VB) and the bottom conduction
band (CB) are mainly composed of Te 5p bonding levels and Cd 5p antibonding levels, respectively.
Below the top VB other VBs are placed, originated from the levels Cd 5s, Cd 4d, and Te 5s.
It is well documented that LDA predictions of the optical bandgap in II-VI are significantly reduced
compared to the experimental values (see Ref.~\onlinecite{vogel96} and references therein).
We have obtained a direct bandgap $E_g=0.573$ eV at the point $\Gamma$, which is 1.3 eV smaller
than the experimental value.\footnote{Our calculation does not includes
the spin-orbit coupling, which splits the top valence band.
Hence, our bandgap must be compared with the average bandgap
$\frac{2}{3}E_g(\Gamma_{8v}-\Gamma_{1c})+\frac{1}{3} E_g(\Gamma_{7v}-\Gamma_{1c})$.}
 Note that the scale of Fig. \ref{fig:pdoscdte} does not allow one to
see the CB edge in the total DOS.
LDA also predicts the Cd  4d electron bands to occur about 3 eV higher than the experimental values,
but still isolated between Te 5s and Cd 5s bands. This underestimation of the Cd 4d
binding energy is due
to the unphysical electron self-interaction in Cd 4d orbitals,
which is not  correctly compensated by
the LDA exchange-correlation functional.\cite{wei88,vogel96}
GGA calculations suffer the same shortcoming. However,
the topology of the band structure is correctly provided by the LDA calculations.
The bandgap underestimation has two sources. One is the error inherent to the LDA
exchange-correlation potential,
and the other is the discontinuity of the exact DFT
exchange correlation potential upon addition of one electron in insulating
solids.\cite{perdew83,sham83}  Calculations of
the exact DFT exchange-correlation potential have shown that
nearly 80\% of  the gap error in Si, GaAs, and AlAs is inherent to exact DFT.\cite{godby88}
However, the calculation of GaAs in Ref.~\onlinecite{godby88}, which is the most similar case to CdTe,
placed the Ga 3d electrons in the Ga frozen core.
When the cation d orbitals are treated self-consistently,
the unphysical proximity of cation d and anion p bands caused by the self-interaction error
repels upwards the higher valence bands, producing a further decrease of the bandgap.

\begin{figure}[!tb]
\includegraphics[width=8.0cm]{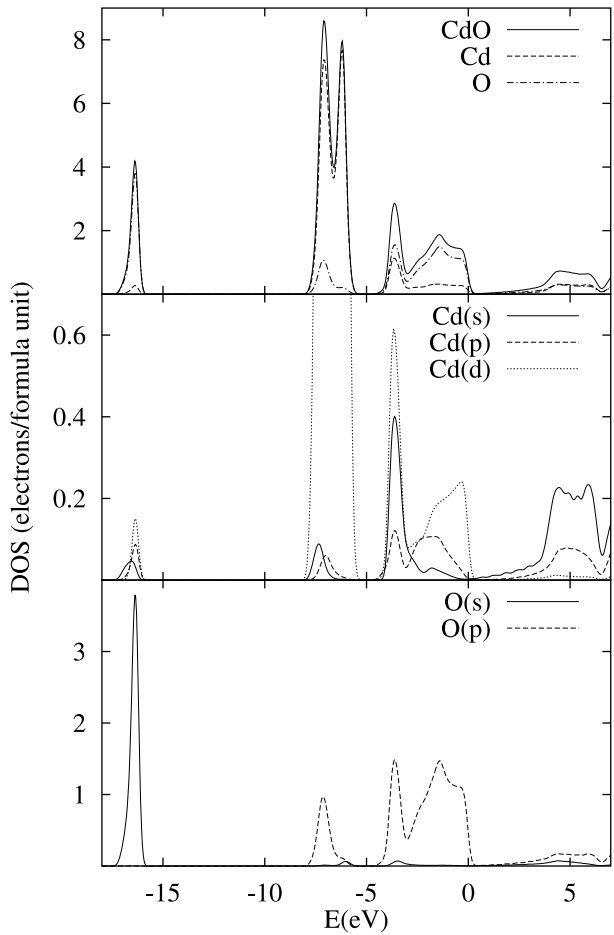}
\caption{Total and partial density of states of CdO. The zero of
the energy scale is set at the valence band maximum at point L. \label{fig:pdoscdo}}
\end{figure}

\subsection{CdO}

\begin{figure}[!tbp]
\includegraphics[width=8.5cm]{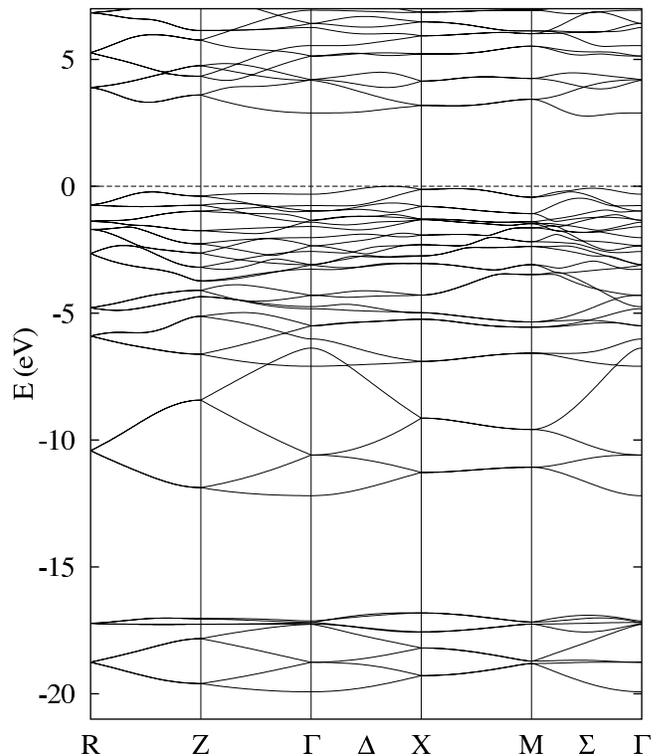}
\caption{Bandstructure of $\alpha$-TeO$_2$. The fractional coordinates of the high symmetry points
of the first Brillouin zone are:
$\Gamma$(0,0,0), M(0.5,0.5,0), Z(0,0,0.5), R(0,0.5,0.5) and X(0.5,0,0).
\label{fig:bandsteo2}}
\end{figure}

CdO is also a II-VI semiconductor. It has  rocksalt structure, where the cations are placed in
the octahedral interstitials of the close packed anions. As oxygen is more electronegative than tellurium,
the Cd-O bonds are more ionic than the Cd-Te ones.
The band structure that we have obtained is similar to others obtained within
standard LDA\cite{vogel96} and GGA.\cite{guerrero02} Hence, we do not repeat it here.
It shows an indirect bandgap, with the valence band maximum (VBM)
at point L, and the conduction band minimum (CBM) at point
$\Gamma$.  LDA and GGA erroneously predict  a
negative bandgap because it improperly accounts for
the electron self-interaction in Cd 4d orbitals.\cite{vogel96,guerrero02} The unphysical
self-interaction in the Cd 4d orbitals raises these levels with respect to the experimental values. The
Cd 4d levels repel the upper O 2p levels,\cite{wei88} thus decreasing the bandgap.
This is the same effect that reduces the
 bandgap in CdTe and other IIB-VIA compounds.
However, the structure of the VB here obtained  is similar to that calculated using
self-interaction- and relaxation-corrected pseudopotentials,\cite{vogel96} where the
gap problem is corrected and the semiconducting character of CdO is restored.

As can be seen from Fig.~\ref{fig:pdoscdo}, the Cd 4d bands are isolated from the top
VBs, which in turn are originated
from the O 2p levels. Nevertheless, as shown by the partial DOS,
there is a certain  mixing between O 2p and Cd 4d levels.
The mixing of Cd 4d and O 2p levels is the origin of the splitting of the Cd 4d band and
the peak at -3.6 eV. \footnote{The partial O 2p character in the Cd 4d band
has been proven experimentally in Ref.~\protect\onlinecite{smith02} using soft X-ray emission
 from the O 1s edge.}
At some points in the Brillouin zone, certain Cd  4d levels have the same
symmetry as the O 2p levels
and form bonding and antibonding combinations, raising the O 2p levels and lowering the Cd 4d
ones.\cite{wei88} Other Cd 4d levels, at the same or at a different $\mathbf{k}$-points,
have different symmetry than O 2p levels, and originate the Cd 4d band at
higher energy. The DOS shows this effect integrated over the Brillouin zone.

\subsection{TeO$_2$}

 \begin{figure}[tbp!]
\includegraphics[width=8.0cm]{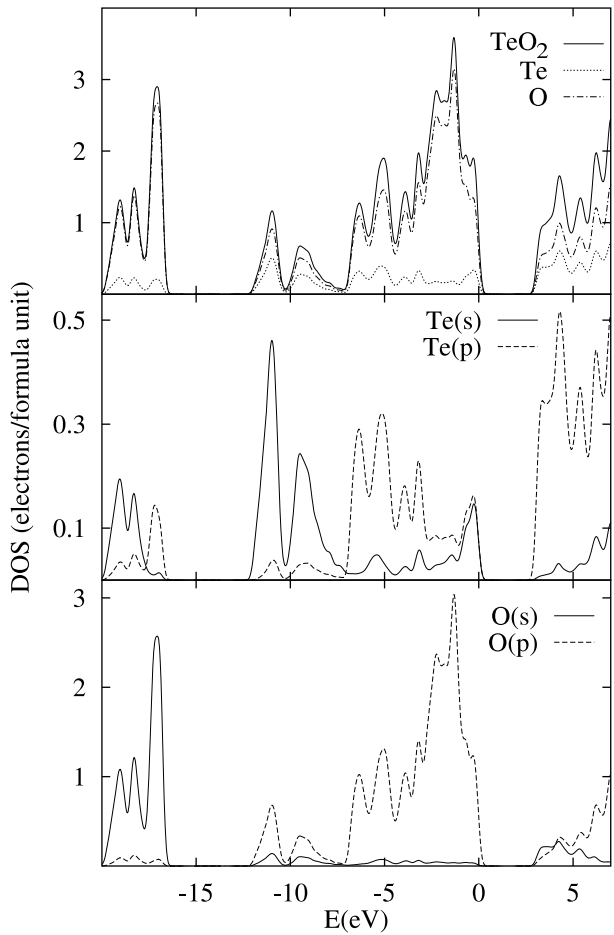}
\caption{Total and partial density of states  in $\alpha$-TeO$_2$.\label{fig:pdosteo2}}
\end{figure}

\begin{figure}[!htb]
\includegraphics[width=4.5cm]{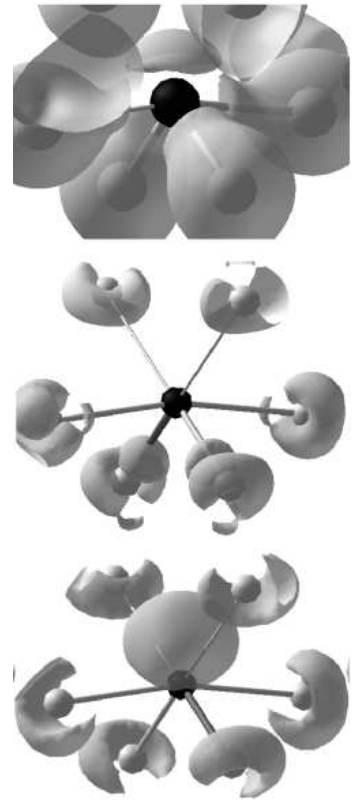}
\caption{Top:  Iso-surface of the electron density at a value of
$\rho=0.51$~\AA$^{-3}$ around
 Te in $\alpha$-TeO$_2$.
Middle: Iso-surface of crystal-minus-atomic difference of electron density at a
value $\Delta\rho=0.16$~\AA$^{-3}$.
Bottom: Iso-surface of the electron localization function at value ELF=0.83. \label{fig:teo4tbp}}
\end{figure}

The band structure and the partial DOS of $\alpha$-TeO$_2$
 are shown in
Figs.~\ref{fig:bandsteo2} and \ref{fig:pdosteo2}.
We have obtained an indirect bandgap of 2.768 eV,
 with the  VBM located at $\Delta$ points
 $\mathbf{k}=(\pm 0.36,0,0),(0,\pm 0.36,0)$ (in fractional units) and the CBM at
 $\Sigma$ points $\mathbf{k}=(\pm 0.25,\pm 0.25,0)$, $(\mp 0.25,\pm 0.25,0)$.
 However,
 the top VB and the bottom CB are very flat and the direct bandgap
 $E_{\Sigma-\Sigma}$=2.851 eV,  at $\mathbf{k}=(\pm 0.2375,\pm 0.2375,0)$
 is very close to the indirect bandgap.
 For practical purposes  $\alpha$-TeO$_2$ can be considered a direct
bandgap material.  The flatness of the band edges also indicates a high degree of  electron
localization.
 The partial DOS shows that contributions from O 2s and 2p, and Te 5s and 5p orbitals are
 spread along a wide range of energies, thus indicating that partially covalent bonds are formed.
Also, the Te atom contributions to the DOS are one order of magnitude smaller than O atom contributions,
 indicating a charge transfer from Te atom to O atom.

 Figure \ref{fig:teo4tbp} shows an iso-surface of the
electronic density and the ELF
around a TeO$_6$ cluster inside the crystal
structure of  $\alpha$-TeO$_2$. The central Te and the four O atoms below it
form the above mentioned trigonal bipyramid.
Several features should be noted in this figure:  (1)
directionality of the Te-O bonds in the tbp; (2) a lone pair over the Te atom,
opposite to the tbp Te-O bonds;
(3) an increment of the charge density  around O atoms in the plane
perpendicular to the O-Te bonds (this feature confirms
the results of cluster calculations\cite{suehara95,suehara98});
(4) there are two regions of high electron localization:  one over the Te atom,
which corresponds to the electrons lone pair, and the other around O atoms
in the plane perpendicular to the O-Te bonds.
Iso-surfaces for higher ELF (not shown in the figure) concentrate at
the Te lone pair.
A smaller increment of the charge density
 ($\Delta\rho\sim 0.05$ \AA$^{-3}$) was also found at the Te lone pair region.

\subsection{CdTeO$_3$}

\begin{figure}[!tbh]
\includegraphics[width=8.5cm]{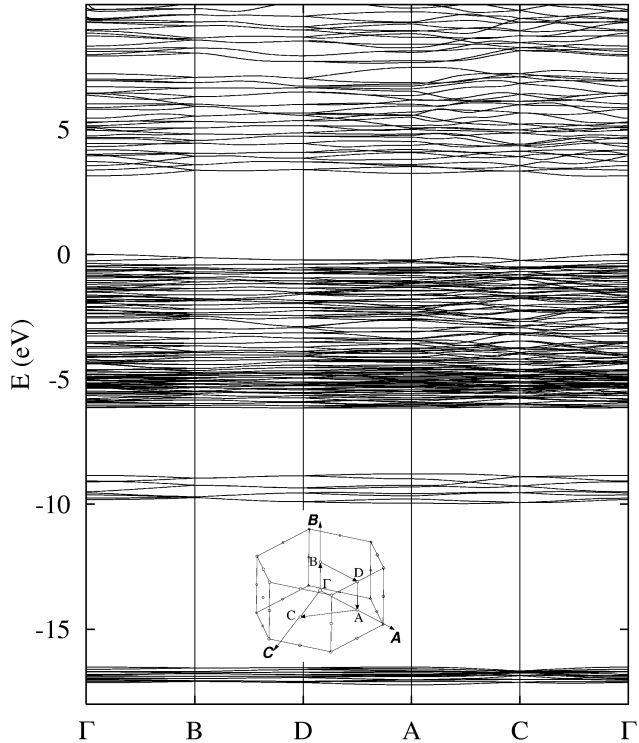}
\caption{Band structure of CdTeO$_3$.
The fractional coordinates of the high symmetry points
of the first Brillouin zone are: $\Gamma$(0,0,0), A(0.5,0,0), B(0,0.5,0),C(0,0,0.5),
and D(0.5,0.5,0).
\label{fig:bandscdteo3}}
\end{figure}

\begin{figure}[!tbh]
\includegraphics[width=8.0cm]{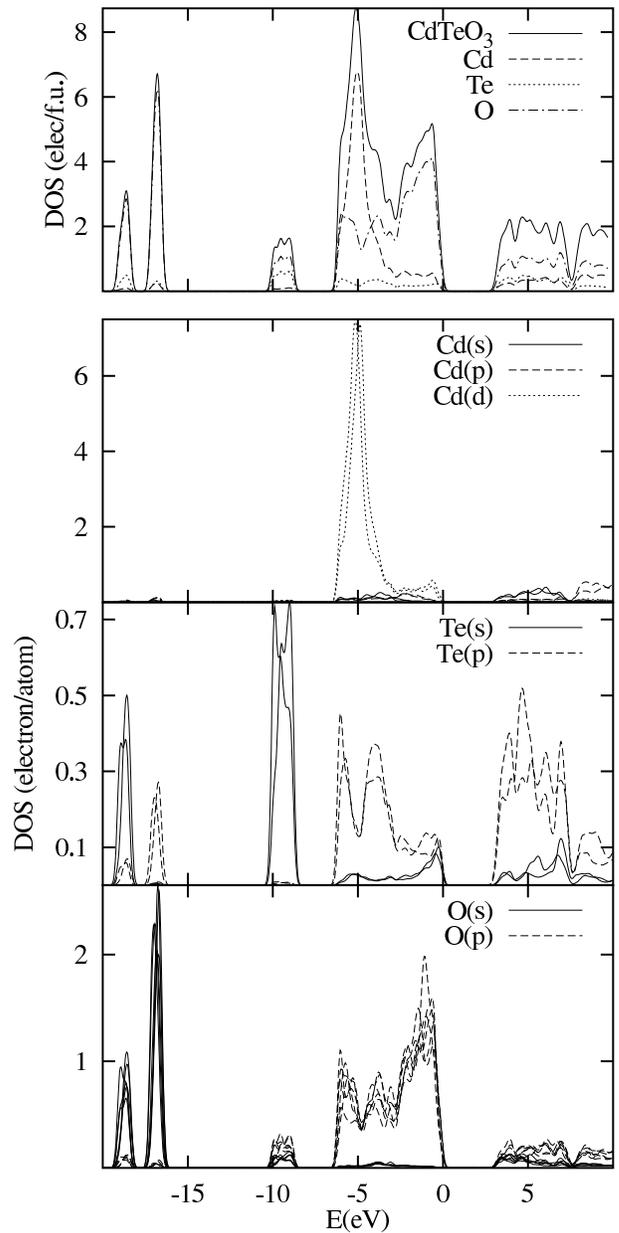}
\caption{Total and partial density of states in CdTeO$_3$
(elec/f.u.=electrons per formula unit). The orbital partial DOS are plotted
for each non-equivalent atom.\label{fig:pdoscdteo3}}
\end{figure}

\begin{figure}[!tbh]
\includegraphics[width=6.0cm]{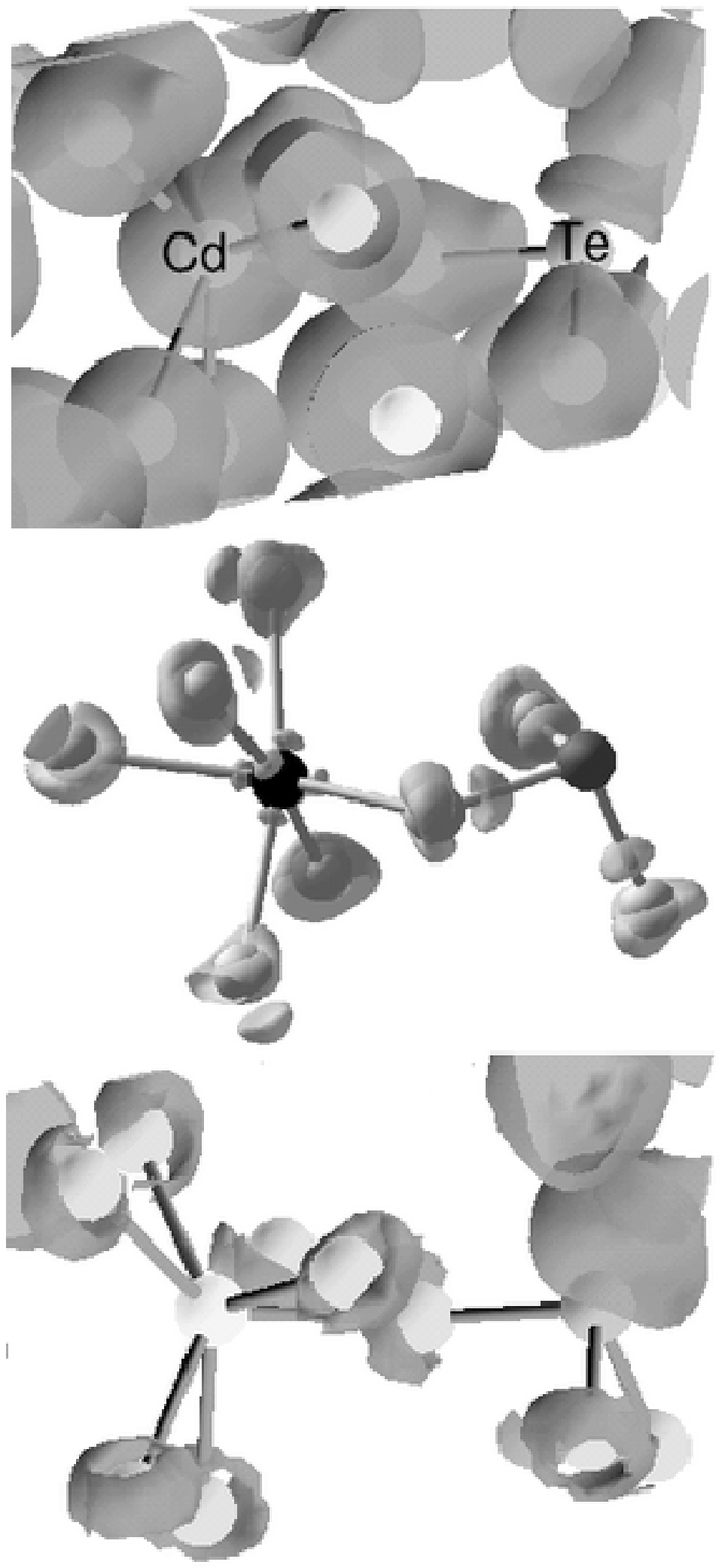}
\caption{Top:  Iso-surface of the electron density at a value of
$\rho=0.51$~\AA$^{-3}$ around
a O$_5$CdOTeO$_2$ cluster inside the CdTeO$_3$ crystal.
Middle: Iso-surface of crystal-minus-atomic difference of electron density at a
value $\Delta\rho=0.16$~\AA$^{-3}$.
Bottom: Iso-surface of the electron localization function at value ELF=0.83.
\label{fig:chgsurf}}
\end{figure}

\begin{figure}[!tbh]
\includegraphics[width=8.5cm]{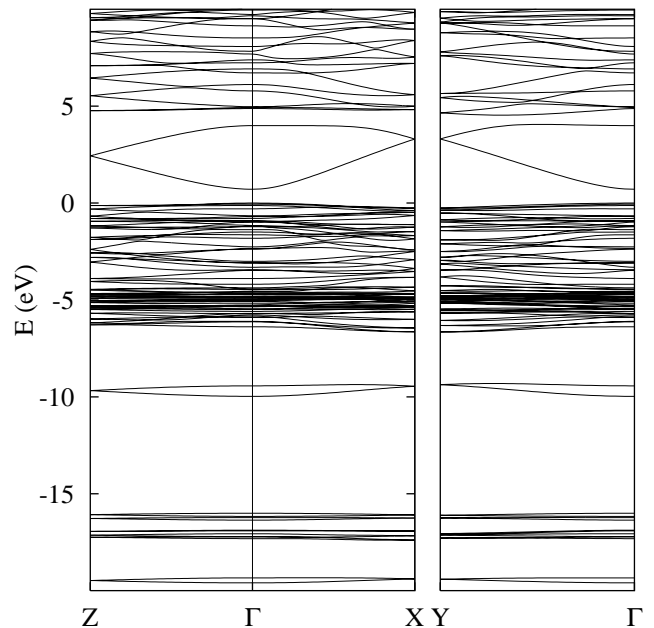}
\caption{Band structure of Cd$_3$TeO$_6$.
The high symmetry points are denoted as in the the orthorhombic system: X(0.5,0,0),
Y(0,0.5,0), Z(0,0,0.5).
\label{fig:bandscd3teo6}}
\end{figure}

Figure \ref{fig:bandscdteo3} shows the band diagram of CdTeO$_3$. The VBM is located at
the $\Gamma$ point, and the CBM is found at
$\mathbf{k}_0=(-3/8,1/8,\pm 3/8)$ and $(3/8,-1/8,\pm 3/8)$ (in fractional units).
Therefore, the bandgap is indirect, with a
value  $E_g$=3.028~eV. However, the direct bandgaps $E_g(\Gamma)=3.142$~eV and
$E_g(\mathbf{k}_0)=3.163$~eV are very close to the indirect gap.
These values are slightly smaller than
the experimental value 3.9 eV reported for amorphous nearly stoichiometric CdTeO$_3$
thin films.\cite{elazhari00}

The total and partial DOS are shown in Fig.~\ref{fig:pdoscdteo3}.
The partial DOS reveals that O 2s orbitals combine with Te 5s and 5p, while O 2p
combine with Te 5s and 5p, and Cd 4d orbitals.
This indicates a complex electronic structure, the main features of
which are: (1) Cd 4d levels concentrate in a single peak in the DOS, with a tail in the region of
O 2p levels, and no gap in between; (2) O 2p levels dominate the top VB
and also contribute to the lower VBs,
being determinant in the O-Te bonds ; (3) Te 5s and Te 5p contributions are distributed throughout
the VB, and are very depopulated in benefit of the O 2p levels; (4) the CB edge is dominated by
O 2p levels, with contributions from Cd 5s and Te 5p levels.
 We conclude that
the Cd-O bonds have a certain covalent character, expressed in the mixing of the 4d levels of Cd with
the 2p levels of O.  The ionic character is suggested by the octahedral coordination of Cd, and by the
spherical symmetry of the charge distribution around Cd atoms (Fig. \ref{fig:chgsurf}).
 However, in contrast to CdO, the partial DOS of CdTeO$_3$  shows that the
Cd 4d levels concentrate in one main peak, which is slightly broader than in CdO, and has a 
background to the higher valence bands without gap.
 This can be understood
as an effect of the more complex atomic environment in this phase with respect to
 rocksalt-structure CdO.
However, the absence of gap between the Cd band and the O 2p band may be
caused by the overall tendency of LDA and GGA to underestimate the binding energies and
the bandwidths.
The Cd 5s levels contribution is present in the conduction band.

Figure \ref{fig:chgsurf} shows an iso-surface of the
electronic density, the charge density difference (crystalline minus atomic),
and the electron localization function (ELF)
around a cluster O$_5$Cd-O-TeO$_2$ inside the crystal
structure of CdTeO$_3$.
Several features should be noted in this image: (1)
spherical symmetry of the electronic cloud around the Cd atom; (2)
directionality of the Te-O bonds; (3) a lone pair near the Te atom
opposite to the Te-O bonds, i.e., over the apex of the pyramid;
(4) an increment of the charge density  around O atoms in the plane
perpendicular to the O-Te bonds (this feature is similar to the case of
$\alpha$-TeO$_2$)
(5) there are two regions of high electron localization:  one over the Te atom,
which corresponds to the electron lone pair, and another one around O atoms
in the plane perpendicular to the O-Te bonds.
Iso-surfaces for higher ELF (not shown in the figure) concentrate at the Te lone pair,
as in paratellurite.  A smaller charge increment was also found at the Te lone pair region.
Note that $\rho=0.51$ \AA$^{-3}$ is a rather
low density (compare with Fig. \ref{fig:denslines}).
For higher values of $\rho$, the iso-surfaces are nearly atom-centered spheres.

\subsection{Cd$_3$TeO$_6$}

\begin{figure}
\includegraphics[width=8.0cm]{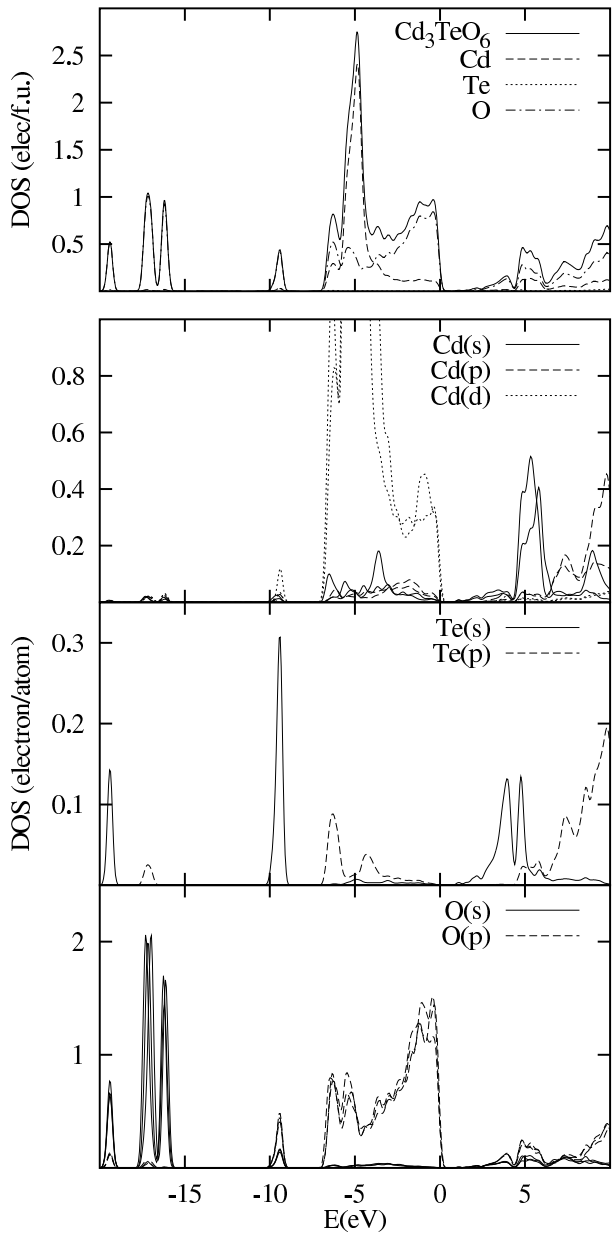}
\caption{Total and partial density of states in Cd$_3$TeO$_6$
 (elec/f.u.=electrons per formula unit).
The partial DOS are plotted  for each non-equivalent atom.
\label{fig:pdoscd3teo6}}
\end{figure}

Figure \ref{fig:bandscd3teo6} shows the band diagram of Cd$_3$TeO$_6$. As can be seen, it is
an insulator material with a direct bandgap $E_g=0.714$ eV at the $\Gamma$ point. The valence bands
are very flat, but the lowest CB shows dispersion and is also isolated from the upper CBs.
From the curvature of the conduction band we have estimated an effective mass
of 0.2 free electron masses, which is comparable to that of II-VI semiconductors.

The total and partial DOS of  Cd$_3$TeO$_6$ are shown in Fig.~\ref{fig:pdoscd3teo6}.
As in CdTeO$_3$, the top VB is built from Cd 4d and O 2p atomic orbitals, and
the lowest VBs are mainly composed of O 2s orbitals with small contributions of
Te 5s and 5p orbitals. The bottom CB has contributions from
all atoms, the largest of which comes from O 2s and Cd 5s.
The most significant difference with CdTeO$_3$
are the triple splitting of the peak associated to O 2s levels
 between $-20$ and $-15$ eV, and the soft
CB edge, which is associated to the small effective mass.

\begin{figure}
\includegraphics[angle=0, width=8.0cm]{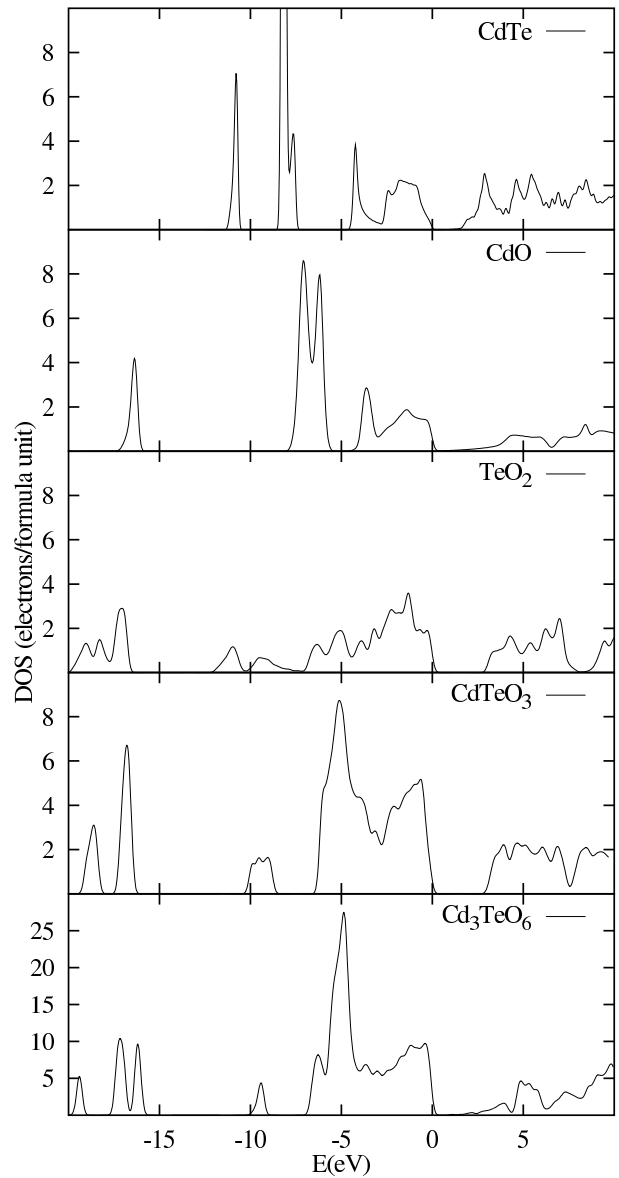}
\caption{Comparative plot of the density of states of CdTe, CdO, TeO$_2$, CdTeO$_3$
 and Cd$_3$TeO$_6$.\label{fig:alldos}}
\end{figure}

We find that the O-Cd bonds are similar to those in the other compounds examined.
Te atoms are
octahedrally coordinated and transfer about 4 electron charges to neighbor O atoms.
The charge density along the line of Te--O bonds is rather similar to the one in CdTeO$_3$,
and there is no electron lone pair bond of Te.
The electron localization function (not shown here) only points to charge localization
around O atoms, which is consistent with the narrow and isolated O 2s bands.

Figure \ref{fig:alldos} shows comparatively the DOS in all the compound studied here.

\section{Discussion}

In order to compare our results for the electronic structure of Cd-Te-O compounds, in
Table \ref{tab:gaps} we show a summary of our calculated bandgaps together with previous results.

The electronic structure of CdO is still not well understood due to both theoretical limitations,
and experimental facts that obscure its intrinsic properties. On one hand,
CdO  presents a large defect concentration, which produces carrier concentrations
as large as 10$^{20}$ cm$^{-3}$,\cite{dou97,wang01cdo}
a nearly metallic conductivity, a Fermi level inside the conduction
band,\cite{dou97,tewari73} a
blocking of the fundamental absorption edge and  a bandgap
shrinkage.\cite{dou97} To
further complicate the interpretation of the optical data, two indirect
bandgaps have been observed below the direct bandgap, which are also
highly sensitive to the temperature. At the
temperature of 100 K the accepted  values \cite{koffyberg76}
for the two indirect bandgaps are 0.84 and 1.09 eV, and 
for the direct bandgap is 2.28 eV.

Previous band diagram calculations
\cite{maschke68,breeze73,tewari73,boettger83,jaffe91,vogel96,dou98,wang01cdo,%
guerrero02} and the present one
provide different interpretations for ordering of the indirect bandgaps.
Early semi-empirical calculations\cite{breeze73,tewari73} and
 correlated Hartree-Fock self-consistent calculations\cite{boettger83,jaffe91} located the
VBM at a $\Sigma$ point, while other semi-empirical,\cite{maschke68}
 LDA,\cite{vogel96,dou98}
periodic Hartree-Fock,\cite{dou98} and screened-exchange LDA\cite{wang01cdo}
calculations, as well as this work, locate the VBM at point L. In all cases the energy
differences between the valence band maxima at L and at $\Sigma$ are small.
Hartree-Fock calculations\cite{boettger83,jaffe91,dou98}
overestimate the bandgaps and widths, while
LDA\cite{vogel96,dou98,wang01cdo} and GGA\cite{guerrero02} underestimate them.
Moreover, the smallest indirect bandgap is negative in some DFT calculations
(Refs. \onlinecite{vogel96,guerrero02} and the present one). Although
the underestimation of
DFT bandgaps is known to be caused by the self-interaction of the
Cd 4d levels and the repulsion from
the higher levels of the $\Gamma$ point, it is not easy to assess the reason for the differences
between the standard-LDA calculations. A common feature of the negative gap calculations is
the use  of pseudopotentials (Refs. \onlinecite{vogel96} and this work) or
supposedly the use of a frozen-core electron configuration.\cite{guerrero02}
The positions of Cd 4d bands have also been a point of interest. While the early
XPS measurement assigned to this band a mean binding energy of 12.8 eV\cite{vesely71}
with respect to the (undefined) Fermi level, recent
measurements have corrected this value to 10.7 eV, or
9.4 eV with respect to the VBM.\cite{dou98}
The best approximation to this experimental energy
up to date is provided
by the periodic HF calculation of Ref.~\onlinecite{dou98}.
LDA calculations
underestimate these binding energies, in this study
giving 6.6 eV.
Our LDA calculation provides a Cd 4d band splitting equal to 0.9 eV,
which agrees with the experimental values of 0.87\cite{vesely71} and 0.8 eV.\cite{scrocco91}

\begin{table*}[!tbh]
\caption{Calculated and experimental values of the fundamental bandgap (in eV)
in all the compounds, and Cd 4d energy in CdO relative to the valence band maximum at point L.
\label{tab:gaps}}
\begin{ruledtabular}
\begin{tabular}{llllllll}
 \multicolumn{8}{l}{CdTe} \\
$E_{g,\Gamma}$    &  Method &  $E_{g,\Gamma}$    & Method  &
$E_{g,\Gamma}$     & Method &&\\
\hline
 & & & &  &  &&\\
1.60 (1.92)   &   TPA\cite{landolt}\footnotemark[1]  &0.3           & PP\cite{vogel96}              & 0.80         &  PP\cite{zakharov94} &&\\
 1.59              &  NLPP\cite{chelikowsky76}                     &0.8           & SIRC-PP\cite{vogel96}   & 1.76         &  PP,GW\cite{zakharov94} & & \\
 1.44              & LAPW\cite{wei88}                                    & 0.47 (0.79)\footnotemark[1] & LMTO-R\cite{cade85} & & & \\
 0.47              & LAPW-SR\cite{wei88}                             &0.51         & LMTO-SR\cite{christensen86} & &&&\\
 0.18              & LAPW-R\cite{wei88}                                &0.29         & LMTO-R\cite{christensen86} & (0.3)  0.573  &PP\footnotemark[2]\footnotemark[3] & &\\
\hline\hline\\
\multicolumn{8}{l}{CdO}  \\
 $E_{g,1}^{ind}$ & $E_{g,2}^{ind}$ & $E_g^{dir}$ & $E$(Cd 4d)& Method  & & & \\
 \hline
  & & & &  &  &&\\
 0.84 & 1.09 & 2.28  &       & TR\cite{koffyberg76} & & &\\
          &          &           &-9.4 & XPS\cite{dou98}          & & &\\
 \\
 $E_{g,L-\Gamma}$ & $E_{g,\Sigma-\Gamma}$ &  $E_{g,\Gamma-\Gamma}$
 &$E$(Cd 4d) & Method & &  & \\
 \hline
 0.8 & 1.2 & 2.38 & -6.1 & APW\cite{maschke68} & & & \\
 1.18 & 1.12 & 2.36 & -5.7 &LCAO\cite{breeze73} & & &\\
 1.11 & 0.95&         & -4.5  & APW\cite{tewari73} &  & & \\
 6.07 &    & 6.56     &  -12.1 & C-HF\cite{jaffe91} & &&\\
  10.9 &   & 11.0  & -13.2 & AE-HF\cite{boettger83} & & & \\
 0.5 & 0.4 & 0.8   & -14.4  &C-HF\cite{boettger83} & & & \\
  -0.6 &     &                &  -6.2   & PP\cite{vogel96} &  & & \\
   1.7  & 1.7   &   3.4         &   -8.2   &SIRC-PP\cite{vogel96} & & &\\
 10.41 &10.44 & 11.66 & -10.3  &AE-HF\cite{dou98} &  & &\\
 0.39 & 0.47 & 1.61    &   -7.1     & AE-LDA\cite{dou98} & & &\\
 $0$ & $0.2$ & 2.36 & & sx-LDA\cite{wang01cdo} &  &&\\
 -0.5 &  & 0.7  &  -6.5 &LAPW-GGA\cite{guerrero02} & & &  \\
-0.465 & -0.386 & 0.712   & -6.6 & PP\footnotemark[3] &&& \\
\hline\hline\\
\multicolumn{4}{l}{TeO$_2$}  &
 \multicolumn{2}{l}{CdTeO$_3$}  & \multicolumn{2}{l}{Cd$_3$TeO$_6$}\\
$E_g$  & Method & $E_g$  & Method &
$E_g$  & Method & $E_g$ & Method \\
\hline
 & & & &  &  &&\\
3.6-3.9  & OPA\cite{uchida71}   &1.7 ($\Lambda-\Gamma$) & LCAO\cite{robertson79}  &3.9 & OPA\cite{elazhari00}\footnotemark[4]     & $\sim$ 1.7  &  guess\footnotemark[5]\\
4.2 (dir)  & OPA\cite{mansingh88}\footnotemark[4]  &1.9 ($\Gamma-\Gamma$) & LCAO\cite{robertson79} & & & &\\
3.75 (ind) & OPA\cite{mansingh88}\footnotemark[4]  &2.768 ($\Delta-\Sigma$)         &  PP\footnotemark[3] & & & & \\
0                  & LMTO-SR\cite{svane87} &2.846 ($\Sigma-\Sigma$)    &PP\footnotemark[3]   & 3.028     &     PP\footnotemark[3]  & 0.714     &  PP\footnotemark[3] \\
 \end{tabular}
 \end{ruledtabular}
 \begin{flushleft}
TPA: Two-photon absorption at T=4.2 K. \\
OPA: One photon absorption at room temperature.\\
TR: Thermoreflectance at T=100 K. \\
PP: Self-consistent LDA with pseudopotentials, which include scalar relativistic effects. \\
NLPP: Empirical non-local pseudopotentials.\\
LAPW: Self-consistent LDA linearized augmented plane waves.\\
LMTO: Self-consistent LDA linearized muffin-tin orbital.\\
NR: No relativistic, SR: Scalar-relativistic (without spin-orbit coupling), R: Relativistic.\\
AE: All electrons calculation. \\
SIRC-PP: Self-consistent LDA with self-interaction and relaxation corrected pseudopotentials.\\
APW: Non-self-consistent augmented plane waves.\\
LCAO: Empirical linear combination of atomic orbitals.\\
C-HF: All electrons correlated Hartree-Fock.\\
sx-LDA: Screened-exchange LDA.\\
\end{flushleft}

\footnotetext[1]{The value in () is the average of the $\Gamma_7-\Gamma_8$ bands
 split by the spin-orbit coupling.}
\footnotetext[2]{The value in () is the estimated correction due to the spin-orbit coupling.}
\footnotetext[3]{Present calculation.}
\footnotetext[4]{In amorphous thin films.}
\footnotetext[5]{Theoretical value plus 1 eV, following the overall trend.}
\end{table*}

The electronic structure of $\alpha$-TeO$_2$ provided by our calculation is compared
in the following with previous findings.
The band diagram of idealized TeO$_2$ with rutile structure
has been calculated using the empirical tight-binding method\cite{robertson79} and
the scalar-relativistic linear muffin-tin LDA method.\cite{svane87} Both calculations
show important differences in the DOS, particularly the self-consistent one,\cite{svane87}
where a metallic band structure was obtained. We performed a test calculation
with the structure of rutile and we also found a metallic DOS. The failure of the
rutile model of TeO$_2$ is due to the asymmetry of the Te-O bonds in the first
coordination shell of Te in the paratellurite structure, as has been demonstrated
in clusters calculations\cite{suehara94,suehara95,suehara98} and in our results.
An XPS spectra was reported and interpreted with the help of a self-consistent-charge
discrete variational X$\alpha$ calculation of TeO$_6$ clusters.\cite{suehara94} Our
findings of the crystal electronic structure calculation support this interpretation and our
DOS corresponds fairly well with the discrete levels of the clusters. The
most  significant difference is the existence of a double peak centered
at $-11$ eV and $-9$ eV in our DOS,
which is composed of  mixed  Te 5s and O 2p orbitals. This
feature corresponds to the band C of Ref. \onlinecite{suehara94}, which
 is not split in the calculations nor in the observed XPS spectrum.
Berthereau \textit{et al}\cite{berthereau96} have
reported Hartree-Fock calculations of charged clusters TeO$_4^{4-}$, which predict the
higher occupied molecular orbital  (HOMO)
to be antibonding and composed of mixed Te~5s-O 2p orbitals and have suggested that
this molecular orbital corresponds to the Te lone pair.
This level can be correlated to the conduction band edge predicted by our crystal calculation.
The crystal calculation is more consistent with less-charged
clusters, where the HOMO of Berthereaud \textit{et al} becomes
the lowest unoccupied molecular orbital, and the HOMO is composed
of non-bonding O 2p levels,\cite{berthereau96} in correspondence with
 our valence band edge.
Based on their study of charged TeO$_6^{n-}$ clusters,
Suehara \textit{et al}\cite{suehara95} estimated the net charge $n$
of the TeO$_6$ structural unit to be $6-$, and the net charge of  Te atoms to be $3+$,
This coincides closely with our estimation based on ab-initio atomic radii (see Appendix).
The optical bandgap provided by our LDA calculation near 2.8 eV can be
compared with available  data.
Uchida\cite{uchida71} measured the absorption in the pre-edge region, in which an Urbach
tail behavior can be observed. Although no attempt was made to estimate a direct or indirect gap, it
is clear from their spectra that the fundamental edge of paratellurite lies in the region
3.6-3.9 eV. Mansingh and Kumar\cite{mansingh88} measured the
optical properties of TeO$_2$ thin films prepared
by vacuum evaporation and found an indirect gap at 3.75 eV and a direct gap at 4.2 eV
related to a strong absorption. Although it is not clear if their as-grown
films were amorphous or polycrystalline, these values provide an indication
of the bandgap for ideal crystals. Hence, the LDA
calculation underestimates the paratellurite bandgap by approximately 1 eV.

The theoretical electronic properties of CdTeO$_3$ and Cd$_3$TeO$_6$ can be discussed only
with reference to amorphous thin films or oxidized CdTe crystals, as there is
no data for crystalline CdTeO$_3$ and Cd$_3$TeO$_6$. The bandgap of
nearly stoichiometric amorphous CdTeO$_3$ is 3.9~eV.\cite{elazhari00}
Hence, LDA underestimates the bandgap in
0.9~eV, which is the same trend as in TeO$_2$, CdO, and CdTe. Based on this trend we
predict a bandgap around 1.7~eV for Cd$_3$TeO$_6$.  This material has recently been
found to be semiconducting and susceptible to n-doping.\cite{shan02}
These experimental findings are in agreement with our predicted gap, conduction band
and effective mass.

Our final comment is devoted to the spin-orbit coupling, which is known to have
a significant effect on the bandgaps of M-Te compounds (M=Zn,Cd,Hg)
(see e.g. Ref.~\onlinecite{wei88}). The spin-orbit splitting of the fundamental bandgap
of these compounds  arises mostly from the Te 5p orbitals.
As our calculation does not include this effect,
our calculated gaps should be regarded as averaged values. However, in the compounds other
than CdTe the  partial DOS of Te 5p levels is largely reduced, and the
O 2s and O 2p orbitals give the greatest contributions to the DOS.
Hence, we expect that the spin-orbit effects on the bandgap should be much smaller than in CdTe.
However, calculations with the  spin-orbit coupling included could be needed
in case that precise measurements of the band properties were produced.

\begin{acknowledgments}
This work has been supported by FONDECYT (Chile) grants  7030083 and 1030063,
Alma Mater grant 32-2000 (Havana University) and a fellowship
from the Third World Academy of Sciences. We also thank
S. Davis for computer assistance.
E. M-P also thanks P. Ordejon, E. Artacho, and A. R. Ruiz-Salvador for many
useful suggestions.
\end{acknowledgments}

\appendix

\section{Determination of the atomic radii}

\begin{figure}[!tbh]
\includegraphics[width=8.5cm]{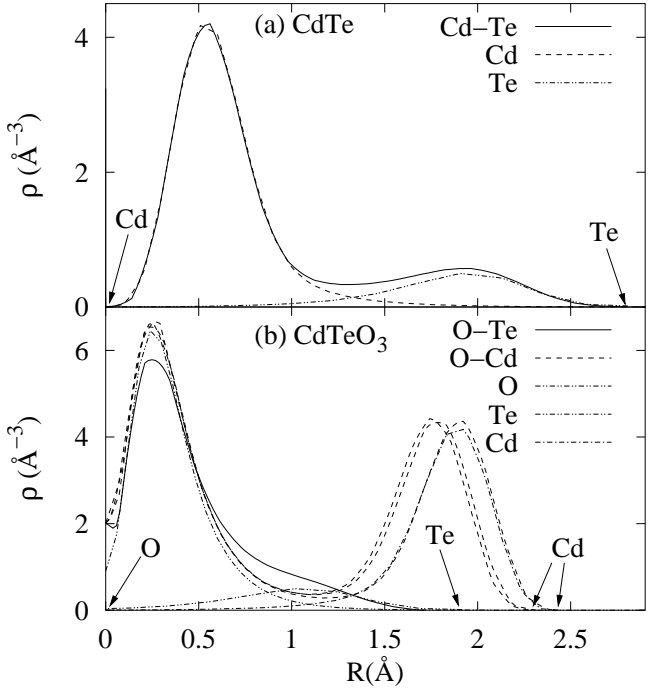}
\caption{Ionic pseudo charges along bonds.\label{fig:denslines}}
\end{figure}

 To obtain the partial DOS it is necessary to define atomic spheres for the integration
 of the local DOS. The atomic radii cannot be unambiguously defined, being
just  measures of their spatial extension, and depend on the environment.
 Partial atomic charges can be obtained  by integration of  the partial atomic DOS
  up to the Fermi level. These partial charges have no exact meaning, but
 can be used as a reference to compare the chemical bonds in the different compounds.

\begin{table}[!t]
\caption{Net atomic charges in CdTeO compounds, determined through a partial
DOS analysis.
\label{tab:chg}}
\begin{ruledtabular}
\begin{tabular}{rcccccc}
  & \multicolumn{2}{c}{Cd} &\multicolumn{2}{c}{Te}&\multicolumn{2}{c}{O}\\
 Compound &  $R$(\AA) &$q$    & $R$(\AA) & q       & $R$(\AA)    & q    \\
 \hline
CdTe (C)  &    1.48   &$-0.46$  & 1.35    & +0.46    &            &       \\
     (A)  &    1.29   &+0.46  & 1.52    & $-0.46$    &            &       \\
CdO  (I)  &    1.09   &+1.6   &         &          &    1.28    & $-1.6$  \\
       (A)   &  1.20    &  +1.1  &         &          &   1.14     &  $-1.1$  \\
TeO$_2$(I)&           &       & 0.80    &  +4.4    &    1.24    &  $-2.2$ \\
       (A)&           &       & 0.9     &  +3.1    &    1.10    &  $-1.55$ \\
CdTeO$_3$(I) & 1.09   &1.45   & 0.80    &  +4.42   &    1.24    &  $-1.96$ \\
         (C) & 1.48   &$-1.05$  & 1.35    & $-0.65$    &    0.73    &  +0.57  \\
         (A) & 1.23   &+0.7   & 0.9     & +3.3     &    1.10    &  $-1.35$ \\
Cd$_3$TeO$_6$(I)& 1.09  &+1.6   & 0.70    & +5.6     &    1.24    &  $-1.7$  \\
                                (A)& 1.23 & +0.97& 0.90    & +3.95   & 1.1         & $-1.14$
\end{tabular}
(A) Ab-initio. (C) Covalent.\cite{webelements} (I) Ionic.\cite{webelements,shannon76}
\end{ruledtabular}
\end{table}

 Usually, tabulated covalent and ionic radii are
 employed for the partial DOS calculations
 (see Table \ref{tab:chg}).  Additionally, we have obtained ab-initio radii
 analyzing the density of charge along the bonds.
 Figure \ref{fig:denslines} shows the electronic pseudocharge density
along the bond directions for all types of bonds present in the Cd-Te-O
system. It can be noticed the difference between the Cd-O, Cd-Te and Te-O
bonds. While the Cd-Te and Cd-O bonds show clear interatomic regions with minimal
charge density, the Te-O bond shows an intermixing of the atomic electron
bonds.
For Cd-O and Cd-Te bonds,
we estimated the atomic radius as the distance between the nucleus and the
minimum of the valence electron density along the bond.
 In the case of Te-O bonds, the electronic density grows continuously from Te
 to O. Hence we set the Te-O boundary at the point where the second derivatives
 is null. As several distances exist in materials other than CdTe and CdO,
 we averaged the obtained atomic radii. Table \ref{tab:chg}  shows
 the atomic charges assuming these ab-initio radii, as well as ionic and covalent
 radii. As can be seen, with the covalent radii, the atomic charges in CdTe
 and CdTeO$_3$ have wrong sign. This is rather unexpected in CdTe, where
 the bond length is well approximated by the sum of the covalent radii of Cd atom and Te atom.
 Using the ionic radii, the charges in TeO$_2$, CdTeO$_3$, and Cd$_3$TeO$_6$
 are in better agreement from values obtained from a chemical point of view,
  but the charge of Te atom
 is extremely high, and O appears over-ionized in TeO$_2$. In this sense,
 the charges estimated
 with the ab-initio radii seem undoubtedly more reasonable. The change observed in the
 partial DOS calculated with  different atomic radii is fundamentally a scale factor.


\end{document}